\documentclass[useAMS,usenatbib]{mn2e}
\usepackage{mathtools}
\usepackage{graphicx}
\usepackage{color}

\title[Modelling molecular clouds from galactic simulations]{Are turbulent spheres suitable initial conditions for star-forming clouds?}
\author[Ramon Rey-Raposo et al.]{Ramon Rey-Raposo$^{1}$\thanks{E-mail:
rrr@astro.ex.ac.uk;}, Clare Dobbs$^{1}$ and Ana Duarte-Cabral$^{1}$.\\
$^{1}$School of Physics \& Astronomy, University of Exeter, Stocker Road, Exeter EX4 4QL\\}
\begin{document}

\date{September 2014}

\pagerange{\pageref{firstpage}--\pageref{lastpage}} \pubyear{2002}

\maketitle

\label{firstpage}

\begin{abstract}
To date, most numerical simulations of molecular clouds, and star formation within them, assume a uniform density sphere or box with an imposed turbulent velocity field. In this work, we select molecular clouds from galactic scale simulations as initial conditions, increase their resolution, and re-simulate them using the SPH code Gadget2. Our approach provides clouds with morphologies, internal structures, and kinematics that constitute more consistent and realistic initial conditions for simulations of star formation. We perform comparisons between molecular clouds derived from a galactic simulation, and spheres of turbulent gas of similar dimensions, mass and velocity dispersion. We focus on properties of the clouds such as their density, velocity structure and star formation rate. We find that the inherited velocity structure of the galactic clouds has a significant impact on the star formation rate and evolution of the cloud. Our results indicate that, although we can follow the time evolution of star formation in any simulated cloud, capturing the entire history is difficult as we ignore any star formation that might have occurred before initialisation. Overall, the turbulent spheres do not match the complexity of the galactic clouds.
\end{abstract}

\begin{keywords}
 galaxies: star formation -- ISM: clouds  -- hydrodynamics -- turbulence. -- gravitation
\end{keywords}

\section{Introduction}

One of the limitations of simulating star formation in molecular clouds is the choice of initial conditions. If excluding entirely arbitrary conditions, this leaves a limited number of geometries. Many studies assume a uniform sphere \citep[e.g.][]{Bateetal2002,Clark2006, Bate2009, Clark2011, Girichidis2011, Federrath2014}, or a periodic box \citep[e.g.][]{GammieOstriker1996, offner2009, Padoan2011, Federrath2012, Myers2013} as initial setups. Other studies use  colliding flows as an attempt to model the large-scale origin of molecular clouds  \citep[e.g.][]{Heitsch2006, Vazquez-Semadeni2006, Hennebelle2008, Banerjee2009, Ntormousi2011, Clark2012}. \citet{Walch2012} model clouds as fractal structures, although their main focus is on examining the propagation of H{\sc ii} regions into structured clouds \citep[see also][]{Gritschneder2009}. Most simulations adopt an imposed turbulent velocity field to model the dynamics of the inter-stellar medium (ISM).\\\\
With all these approaches, there are concerns about how the initial conditions affect the results, such as the evolution of the cloud, the resulting density and velocity structure and the star formation rate. One alternative way to select initial conditions is to use clouds extracted from  full-scale galaxy simulations. This is the approach we take in this Letter, where we extract clouds from \cite{Dobbs2013a}, and Dobbs 2014 (submitted). Because these clouds in the galactic simulations have a limited number of particles, we resimulate the extracted clouds with higher resolution. We compare our results to the more typical approach of simulating an initially uniform sphere subject to a turbulent velocity field. Our Letter is organised as follows: In Section 2 we briefly explain the details of simulations. In Section 3 we discuss the density structure and star formation rate in both the galactic clouds and the turbulent spheres. Finally, in Section 4 we summarise our main findings.
\begin{figure} 
\centering
\includegraphics[width=95mm]{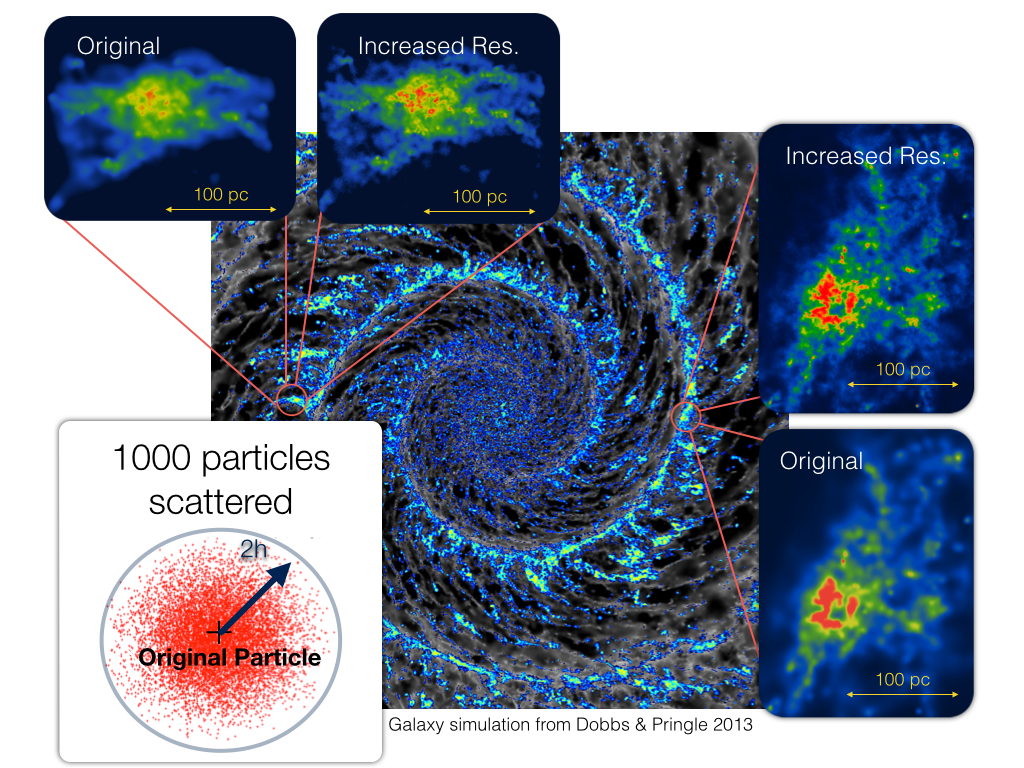}
\caption{Top-down view of the simulated galaxy showing the position of two selected clouds. On the top left, we display the column density plot of an inter-arm cloud (Cloud B), with both the original and increased resolution. On the right, we show the column density plot of an in-arm cloud (Cloud A) at both resolutions. In the lower left we show our method for increasing the resolution, whereby we distribute extra particles according to the SPH kernel of smoothing length h.} 
\label{fig:Fig1}
\end{figure}
\section{Details of simulations}
\begin{table}  

\caption{Mass, radius, velocity dispersion, virial parameter and number of particles of each simulated cloud.}   
\centering 
\label{tab:summary}
\begin{tabular}{c c c c c c c} 

\hline\hline 

Cloud & Mass & R &  $\sigma$ & $\alpha$ & Part $\#$  \\ [0.5ex]

      & $(M_{\odot})$ & (pc) &  (km/s) &    \\   
\hline 

Cloud A & $4.3 \times 10^{6}$ & 100 & 8.75 & 2.07 & $9.6 \times 10^6$\\ 

Sphere A & $3.0 \times 10^{6}$ & 100 & 7.60 & 2.24 &  $1.00 \times 10^7$\\

Cloud B & $2.6 \times 10^{6}$ & 100 & 5.17 & 1.18 & $1.01 \times 10^7$ \\

Sphere B & $3.6 \times 10^{6}$ & 100 & 6.08 & 1.19 & $1.00 \times 10^7$\\

Cloud C & $1.4 \times 10^{6}$ & 100 & 7.80 & 5.02 & $1.09 \times 10^7$\\

Early A & $6.1 \times 10^{6}$ & 200 & 11.48 & 5.01 & $1.07 \times 10^7$\\ [1ex] 

\hline 

\end{tabular} 
\label{table:nonlin} 
\end{table}
Our starting ground is the galaxy simulation described in \cite{Dobbs2013a} and shown in Figure \ref{fig:Fig1}. This simulation includes self gravity, ISM cooling and heating and stellar feedback. The particle mass in the galaxy simulation is 312.5 M$_{\odot}$, and the giant molecular clouds (GMCs) contain $\sim10^4$ particles. We extract these clouds by selecting a box of gas (L $\sim$ 100 pc) which includes the cloud, and increase the resolution by a factor of N. To do so, we split each particle N times, distributing N-1 new particles according to the SPH kernel (as shown on the bottom left box in Figure \ref{fig:Fig1}). The velocities are kept the same as the original particle, to conserve energy and momentum. Although observed clouds are very cold, T $\sim$10 K, we performed isothermal simulations with 50 K which ensures that the Jeans mass is well resolved \citet{Bate1997}, \citep[see also][for more recent studies]{Federrath2011, Federrath2014}. We also ran simulations with 20 K though (not shown), which gave similar overall results.\\\\ 

In Figure \ref{fig:Fig1}, we show the galactic simulation at 250 Myr from which we have selected two clouds, one inside a spiral arm (Cloud A), and the other in an inter-arm region (Cloud B), both with an approximate radius of 100 pc. We show these two clouds in the two onsets of Figure 1, with the original and increased resolution. To compare these models, we have created two turbulent spheres of 100 pc radius (Spheres A and B), with similar virial parameters \citep[as defined in][]{Dobbs2011a} to Clouds A and B ($\alpha \sim 1$ and $\alpha \sim 2$ respectively). The two clouds are both found to exhibit a velocity dispersion relation of $\sigma \propto r^{1/2}$ \citep[in accordance with observed and other simulated clouds, e.g.][]{Roman-Duval2011,Federrath2011}, so we set up the spheres with a velocity power spectrum of P $\propto k^{-4}$ to give a similar scaling relation \citep{Myers&Gammie1999}. The masses and amplitudes of the velocities are scaled to give similar kinetic and gravitational energies and virial parameters to Clouds A and B.\\\\
We take Clouds A and B from a snapshot of the galactic simulation and although the original galactic simulation included the prior evolution of these clouds, it did not contain sink particles, or follow star formation in detail. We traced back the gas in Cloud A to a time of 240 Myr in order to follow the preceeding stages of Cloud A's evolution when the gas was less gravitationally bound (we call this model Early A). Lastly we wanted to test if the method of extracting galactic clouds is robust, given the large increase in resolution. Hence, we have selected a cloud in a spiral arm taken from a simulation by Dobbs 2014 (submitted), which models gas going through a spiral arm with a particle mass of 3.85 M$_{\odot}$ (Cloud C). For Cloud C, we only increase the resolution by a factor of N = 30. The main parameters of all the clouds are summarised in Table \ref{tab:summary}.\\\\
We follow the evolution of the clouds using the SPH code Gadget2 \citep{Springel2005}. Our simulations are isothermal with a temperature of 50K. We include sink particles similar to \citet{Bate1995a} at densities of $\rho_{sink} = 1.6 \times 10^4$ cm$^{-3}$ with a sink radius $R_{sink}$ = 0.1 pc using the implementation in \citet{Clark2008} \citep[based on][]{Jappsen2005}. We run simulations of the GMCs for 16 Myr and the spheres for 24 Myr, which corresponds to at least 3 free fall times for all of the clouds. We do not include the galactic potential in our simulations \citep[as described in][]{Dobbs2006}. We tested its impact on Early A, the biggest cloud, with little effect: the rotational period of the galaxy ($\sim$ 220 Myr) is much greater than the simulation time of our clouds, and the clouds do not traverse between the spiral arms and interarm regions in any of our calculations. The effect of both feedback and cooling are not included but will be investigated in future work.
 
\begin{figure*} 
\centering
\includegraphics[width=190mm]{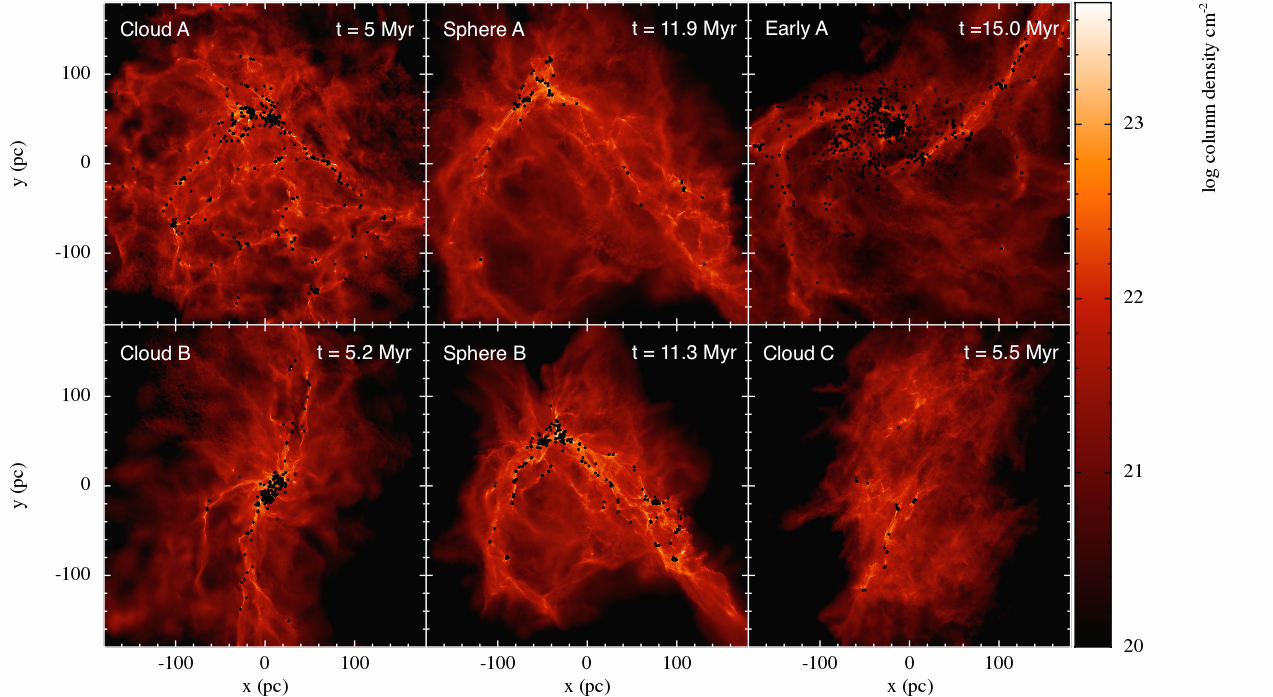}
\caption{Column density plots of the clouds 5 Myr after the first star is formed are shown except for Early A (which is shown at 15 Myr to compare it with Cloud A). The sink particles are represented by black dots. The galactic clouds show a variety of density configurations, with Cloud A showing a rather complex network of filaments, Cloud B and Early A being dominated by one main long dense filament, and Cloud C appearing as rather diffuse and barely substructured cloud.  Sphere A and B are dominated by two dense filaments that coalesce in the centre of the cloud.} 
\label{fig:Fig2}
\end{figure*}
\section{Comparing the evolution of the galactic clouds with the turbulent spheres}

We show the column density plots of the six clouds 5~Myr after the first star is formed in Figure \ref{fig:Fig2} (except for Early A). In the galactic simulation Early A evolves into Cloud A after 10 Myr. Therefore, in Figure \ref{fig:Fig2} we show Early A at 15 Myr, to compare it with Cloud A at 5 Myr. All clouds show a complex filamentary network and are highly structured, whether using the initial conditions from the galaxy, or the turbulent spheres. The main structures in Clouds A and B reflect the galactic structure - the most dominant filaments in each are aligned with a spiral arm and inter-arm spur respectively. If we compare Cloud A and Sphere A we can see that star formation is more widespread in Cloud A, rather than restricted to one or two main filaments, as is the case for Sphere A.  For both Cloud B and Sphere B, the cloud evolution and location of star formation is dominated by one or two long filaments. Both Cloud A and Early A show two main filaments and a cluster of stars in the centre. However, the evolution of Early A is altered by the large number of sink particles formed at early times. Cloud C, our last cloud, has formed far fewer stars and has less dense features compared to all other clouds, even though it starts from a similar unbound state (as e.g. Early A).\\

\begin{figure} 
 \centering
 \includegraphics[width=90mm]{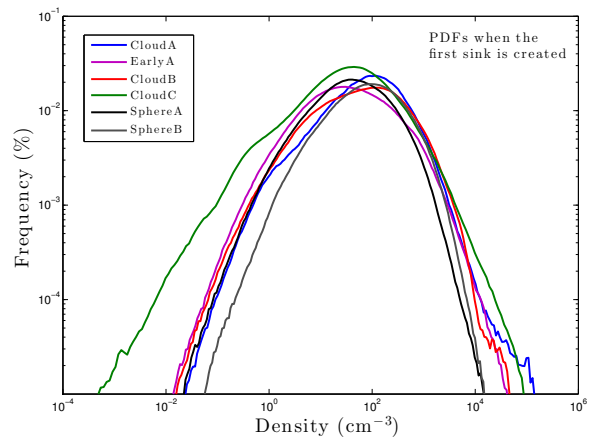}
 \caption{Density PDFs are shown for all the simulated clouds. The clouds have similars PDF compared with the turbulent spheres. Cloud C has more diffuse gas than the rest of the clouds.}
 \label{fig:initialpdf}
 \end{figure}

In Figure \ref{fig:initialpdf} we show the density PDF (Probability Density Function, see   \citet{Vazquez-Semadeni1994, Federrath2008}) for all the clouds. The PDFs show good agreement between the spheres and galactic GMCs. All PDFs are similar, except for Cloud C, which stands out for containing significantly less dense gas when compared to the other clouds. In the beginning of the simulation, the PDFs for the turbulent spheres are obviously narrower in comparison to the rest of the GMCs.\\

\begin{figure} 
\centering
\includegraphics[width=95mm]{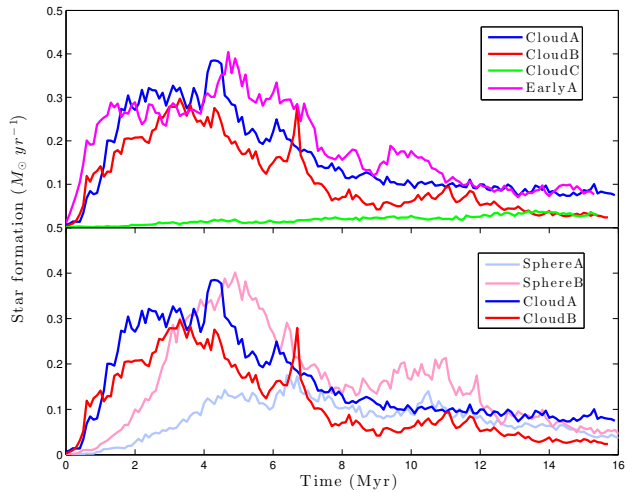}
\caption{On the top panel the SFR for the galactic clouds are shown. Clouds A, B and Early A present a similar behaviour creating stars in the beginning and gradually decreasing afterwards. The efficiency for Cloud C is much lower. On the bottom panel we compare Clouds A and B with the spheres A and B. We have set the origin of time when the first sink is created.}
\label{fig:sfr}
\end{figure}

In Figure \ref{fig:sfr} we show the star formation rate defined as SFR(t) =  $\dot{M}$$_*$(t), where $\dot{M}$$_*$(t) is the time derivative of the mass contained in sinks. We have used a timestep of 0.1 Myr. In the top panel we show the SFR of the galactic clouds. The star formation process is similar for A, B and Early A, starting almost from initialisation, as these clouds already have overdense regions. Once the initial star formation burst is over, the SFR decreases during the remainder of the simulation because there is less gas available (as it has been accreted by the sinks). For Cloud C, the star formation rate behaves differently - it increases slowly, and is significantly lower than the other clouds during most part of the simulation. On the bottom panel we show the SFR for clouds and spheres A and B. We have set the origin of time when the first sink is formed. The spheres need 6 - 7 Myr to create the first sink, and another 4 - 5 Myr to reach the peak of the SFR. At later times the SFRs are very similar for both the GMCs and the spheres. The total star formation efficiencies we obtain for all cases are high ($\sim50 \%$) compared with the observed $\sim5 \%$  \citep[e.g.][]{Bigiel2008,Evans2009}. However, we have not included magnetic fields or feedback processes which likely reduce the efficiencies to similar values of other simulations in the literature $\sim10 - 20 \%$ \citep[e.g.][]{Price2009, Dobbs2011b, Federrath2012,Federrath2013a, Dale2014}. \\

\begin{figure*} 
\centering
\includegraphics[width=190mm]{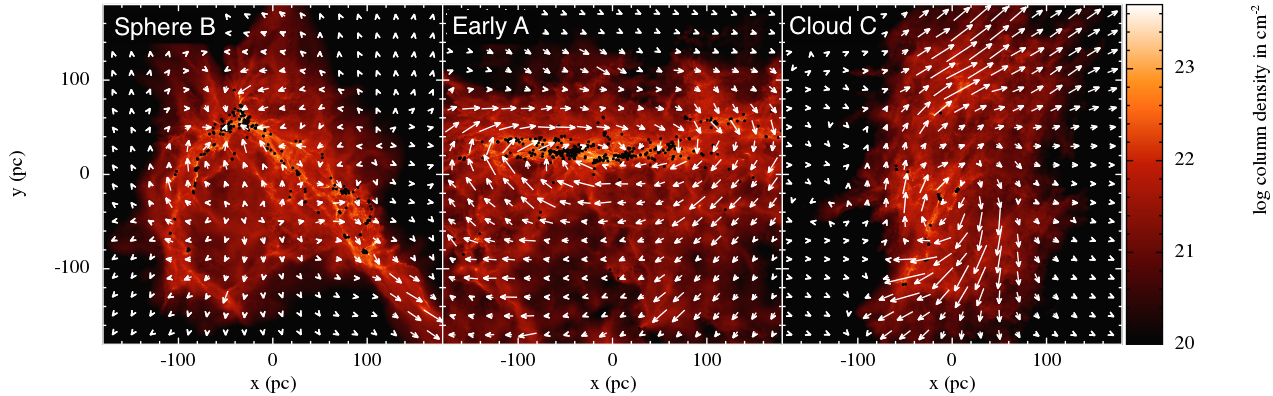}
\caption{Projected velocity field (in white arrows) superposed on the column density maps for three studied clouds 5 Myr, after the first star is formed. In the sphere, the velocity field follows the direction of the gravitational collapse, and the highest velocities are in the vicinity of a filament. For Early A and Cloud C the velocities inherited from the galactic simulations are more important than those arising from the gravitational collapse (except in the densest areas). The shear flows that inhibit star formation in Cloud C are patent.}
\label{fig:vfield}
\end{figure*}

The global evolution of Cloud C is substantially different to the other clouds. We suspected this was a consequence of the large scale velocity field. We include a velocity map of Cloud C, Early A and Sphere B, 5 Myr after the first sink is created in Figure \ref{fig:vfield}. For Sphere B, the velocity field mainly traces the gravitational collapse in the main filaments where star formation happens. The velocity field for Early A shows stronger rotation, but there is still convergence in the centre where stars are forming. Cloud C has also a peculiar velocity field also inherited from the galactic simulation. It has a strong pair of divergent flows in the northern and southern regions, which results in the disruption of the cloud, inhibiting star formation. To check whether the difference between Cloud C and the other examples was linked to how much we increase the resolution, we also selected another cloud from the spiral arm simulation of Dobbs 2014 (submitted). The SFR in this last example (not shown in Figure \ref{fig:sfr}) was higher, and comparable to the other simulations. This confirms that the shear flows in Figure \ref{fig:vfield} are responsible for the difference in star formation rate for Cloud C. The effects of the different velocity fields are clearer when visualising the evolution of the clouds and spheres in a movie.

\section{Conclusions}
In this letter we performed numerical simulations of clouds that have been extracted from galactic simulations. We selected four clouds and modelled two turbulent spheres that resemble two of the galactic clouds. We explored the differences and similarities of using turbulent spheres and GMCs as initial conditions to model the star formation process. The main advantage of the GMCs compared to the turbulent spheres is that they provide a wider variety of morphologies and velocity structures which influence the clouds' evolution and properties.\\

There are some clear similarities between the simulated GMCs and turbulent spheres, namely their PDFs and star formation rates or efficiencies. Although the initial PDFs of the spheres are narrower, they eventually become comparable to most of the GMCs at late times. The spheres also have comparable SFRs once they have evolved and formed dense areas able to produce stars. However the GMCs can evolve to show quite different behaviour from each other, and the spheres, dependent on their initial conditions. The velocity field from larger (galactic) scales affects the morphology, kinematics and can effect the star formation in those clouds. The influence of the inherited properties appears to have a greater impact on star formation than the virial parameter of the clouds. For instance, Cloud C and Early A have similar virial parameters, but the star formation rate of Early A is more comparable to the other models, whereas in Cloud C it is inhibited by the inherited shear flows. In essence, the spheres tend to be dominated by gravitational infall, whereas for the GMCs the large scale velocity field can be equally important. Our conclusions are in agreement with \citet{Federrath2012}. They find that the compressive and solenoidal components of a turbulent velocity field (quantified by the mode mixture parameter $b$) have a large impact on star formation. This constitutes the main advantage of the GMCs, as creating such different environments which would be difficult to reproduce with turbulent spheres.\\ 

Another advantage with respect to turbulent spheres is that as well as modelling clouds in different environments (for example arm and inter-arm regions), we can also study different stages of their evolution. We found that Cloud A and Early A, which should be the same cloud, have different morphologies due to following sink particle creation in Early A. Our results are somewhat extreme, as we do not include feedback and the star formation rate in Early A is far too high. However this highlights that likely all simulations of isolated clouds will miss a previous star formation history. This problem can perhaps be lessened when using galactic simulations and tracing clouds back to earlier stages.

\section{Acknowledgements}
We thank an anonymous referee for suggestions which helped improve the paper, and Paul Clark for comments on an earlier draft. The calculations for this paper were performed on the University of Exeter Supercomputer, a DiRAC Facility jointly funded by STFC, the Large Facilities Capital Fund of BIS, and the University of Exeter. RRR, CLD and ADC acknowledge funding from the European Research Council for the FP7 ERC starting grant project LOCALSTAR. Fig. \ref{fig:Fig1}, Fig. \ref{fig:Fig2} and Fig. \ref{fig:vfield} were produced using SPLASH \citep{Price2007}. 
\bibliographystyle{mn2e}

\bibliography{rrr_MNRAS_Letter}
\label{lastpage}

\end{document}